%Paper: 9109024
%From: aoki@madonna.physics.ucla.edu (Ken-ichiro Aoki)
%Date: Thu, 12 Sep 91 19:38:05 PDT
%Date (revised): Tue, 24 Sep 91 20:48:39 PDT

\input phyzzx
\def\ucla{Department of Physics\break
      University of California Los Angeles\break
        Los Angeles, California 90024--1547}
\def\nl{\hfil\break}

\def\anp{{ \sl Ann. Phys. }}

\def\ijp{{ \sl Int. J. of Mod. Phys. }}

\def\mpl{{ \sl Mod. Phys. Lett. }}

\def\npb{{ \sl Nucl. Phys. }}

\def\prd{{ \sl Phys. Rev. }}
\def\prl{{ \sl Phys. Rev. Lett. }}
\def\plb{{ \sl Phys. Lett. }}

\def\Re{{\rm Re\,}}

\def\undertext#1{\vtop{\hbox{#1}\kern 1pt \hrule}}
\def\half{{1\over2}}
\def\c#1{{\cal{#1}}}
\def\slash#1{\hbox{{$#1$}\kern-0.5em\raise-0.1ex\hbox{/}}}
\def\g{{\hat g}}
\def\gm{{\hat g_{mn}}}
\def\half{{1 \over 2}}
\font\cmss=cmss10 \font\cmsss=cmss10 at 7pt
\def\IZ{\relax\ifmmode\mathchoice
{\hbox{\cmss Z\kern-.4em Z}}{\hbox{\cmss Z\kern-.4em Z}}
{\lower.9pt\hbox{\cmsss Z\kern-.4em Z}}
{\lower1.2pt\hbox{\cmsss Z\kern-.4em Z}}\else{\cmss Z\kern-.4em Z}\fi}
\overfullrule=0pt

\overfullrule=0pt
\Pubnum={UCLA/91/TEP/32}
\date={August, 1991}
\titlepage
\title{On the Liouville Approach to Correlation Functions for 2-D Quantum
Gravity
\foot{\rm Work supported in part by the National Science Foundation grant
PHY--89--15286.} }
\author{\rm Kenichiro Aoki and Eric D'Hoker\foot{\rm
email:~aoki@uclahep.bitnet, dhoker@uclahep.bitnet} }
\address{\ucla}
\abstract{We evaluate the three point function for arbitrary states
in bosonic minimal models on the sphere coupled
to quantum gravity in two dimensions.
The validity of the formal continuation in the number of
Liouville screening charge insertions is shown directly
from the Liouville functional integral using semi-classical methods.}
\medskip
\endpage
\def\al#1{\alpha_{#1}}
\def\alr#1{\alpha_{r'_{#1}r_{#1}}}
\def\bestar{\alpha}
\def\frac#1{\left\langle{#1}\right\rangle}
\def\corr#1{\left\langle{#1}\right\rangle}
\def\prodone#1{\prod_{j=1}^{#1}}

\def\op{\psi}
\def\opcft#1{\psi^{\rm matter} _{r'_{#1}r_{#1}}(z_{#1})}
\def\d{\partial}
\def\inv#1{{#1}^{-1}\!}
\def\beh#1{\beta(h_{#1})}

\def\fpi{{1 \over 4 \pi}}
\REF\POL{A.M. Polyakov, \plb B103 (1981) 207}
\REF\CDG{T. Curtright and C. Thorn, \prl 48 (1982) 1309 \nl
	E. D'Hoker and R. Jackiw, \prd D26 (1982) 3517 \nl
	J.-L. Gervais and A. Neveu, \npb B209 (1982) 125; B224 (1983) 329}
\REF\DDK{F. David and E. Guitter, Euro. Phys. Lett. 3 (1987) 1169 \nl
	F. David, \mpl A3 (1988) 1651\nl
	J. Distler and H. Kawai, \npb B321 (1988) 509}
\REF\KPZ{A. M. Polyakov, \mpl A2 (1987) 893 \nl
	V. Knizhnik, A. M. Polyakov and A. Zamolodchikov, \mpl A3 (1988) 819}
\REF\BDS{E. Br\'ezin and V. Kazakov, \plb B236 (1990) 144 \nl
	M. Douglas and S. Shenker, \npb B335 (1990) 635 \nl
	D. J. Gross, A. A. Migdal, \prl 64 (1990) 127 }
\REF\DK{P. Di Francesco, D. Kutasov, \npb B342 (1990) 475}
\REF\TOP{E. Witten, \npb B340 (1990) 280 \nl
	J. Distler, \npb B342 (1990) 523 \nl
	R. Dijkgraaf and E. Witten, \npb B342 (1990) 486 \nl
	E. Verlinde and H. Verlinde, IASSNS-HEP-90/40 (1990) preprint \nl
	M. Fukuma, H. Kawai, R. Nakayama, \ijp{\bf A6} (1991) 1385\nl
	R. Dijkgraaf, E. Verlinde and H. Verlinde, PUPT-1184 (1990) preprint\nl
	J. Distler and P. Nelson, UPR-0438T, PUPT-1201 (1990) preprint}
\REF\BDG{E. Braaten, T. Curtright, G. Ghandour and C. Thorn,
	\anp (NY) 147 (1983) 365; 153 (1984) \nl
	J.-L. Gervais and A. Neveu, \npb B257 (1985) 59; B264 (1986) 557}
\REF\PO{J. Polchinski, Texas preprint UTTG-1990 (1990) \nl
	N. Seiberg, Rutgers preprint RU-90-29}
\REF\GL{M. Goulian, M. Li, \prl 66 (1991) 2051}
\REF\KIT{Y. Kitazawa, HUTP-91/A013 (1991) preprint}
\REF\DOT{Vl. S. Dotsenko, PAR-LPTHE 91-18 (1991) preprint}
\REF\DFK{P. Di Francesco and D. Kutasov, Phys. Lett. B261 (1991) 385\nl
	J.G. Russo, SU-ITP-895 (1991) preprint}
\REF\BPZ{A. Belavin, A. M. Polyakov and A. Zamolodchikov,
	\npb B241 (1984) 333}
\REF\SUP{K. Aoki and E. D'Hoker, UCLA/91/TEP/33 (1991) preprint}
\REF\MDP{N. Mavromatos and J. Miramontes, \mpl A4 (1989) 1849 \nl
	E. D'Hoker and P. S. Kurzepa, \mpl A5 (1990) 1411}
\REF\EDH{E. D'Hoker, \mpl A6 (1991) 745}
\REF\DF{Vl. S. Dotsenko and V. A.  Fateev, \npb B240 (1984) 312;
	B251 (1985) 691}
\REF\FQS{D. Friedan, Z, Qiu and S. Shenker, \prl 52 (1984) 1575}
\REF\EFJ{E. D'Hoker, D. Z. Freedman and R. Jackiw, \prd D28 (1983) 2583}
\REF\SZE{G. Szeg\"o, ``Orthogonal Polynomials"  Am. Math. Soc. XXIII, 1959}
\REF\BOAS{R. Boas, {\sl ``Entire Functions"}, Academic Press, 1954}
\lettersize\singlespace
\chapter{Introduction}
The Liouville field theory approach to 2-D quantum gravity
and non-critical string theory arises naturally
from the geometry of the 2-D worldsheet [\POL].
Yet, despite the classical integrability of the theory,
its quantization is still incompletely understood.
Progress was made by assuming free field operator product
expansions to compute the scaling dimensions of exponential
operators [\CDG,\DDK].
The results agree with the worldsheet lightcone gauge
approach [\KPZ], with matrix models [\BDS,\DK]
and topological field theory [\TOP].
Still, Liouville dynamics is not completely equivalent
to that of free field theory in general [\CDG,\BDG,\PO].
More recently, it was proposed to push the free field
approach yet one step further, and to compute correlation
functions as in free field theory but with the Liouville
exponential interaction interpreted as a
screening charge [\GL,\KIT,\DOT,\DFK].
The three point function obtained in this way
agrees with matrix model predictions.

In the present paper, we generalize calculations
of [\GL,\KIT]
to the case of arbitrary external states
in the Kac table of bosonic minimal models [\BPZ].
The integration over the constant Liouville mode, which yields
the screening charge interpretation to the Liouville
interaction, gives rise to a single screening insertion
in the Liouville sector.
This is to be contrasted with the approach of [\DOT]
where the presence of the two Liouville screening charges is
postulated.
Still, our final results agree with those of [\DOT]
and also with matrix model calculations.

The number of screening charges ($s$) brought down by
integrating out the constant Liouville mode is not
in general an integer in this approach.
In [\GL] it was proposed to rearrange the explicit answer for
integer $s$ in a form that formally makes sense
for all complex $s$.
The justification for this procedure is incomplete.
Two meromorphic functions in $s$ that agree on all
positive integers need not be the same throughout
the complex plane, unless further information
on their behavior at infinity
provided.
We shall justify the continuation procedure here by
showing that the asymptotics for $|s|\rightarrow \infty$ of the
Liouville functional integral and of the explicit answer in terms
of $\Gamma$ functions in fact agree as well.
To do so, we use a semi-classical approximation
to the Liouville functional integral precisely
valid for $|s|\rightarrow \infty$.
This proof completes the justification of the formal
continuation procedure in $s$.

The great advantage of the Liouville approach
to 2-D quantum gravity is that whatever techniques
were used successfully in the bosonic theory can
be readily extended to the super-Liouville case.
This is particularly important since no matrix
or topological field theory formulations for 2-D
quantum supergravity are presently available.
The calculation of the three point function for
super-minimal models is carried out in a companion paper [\SUP].

\chapter{General Three Point Functions in Minimal Models}
We apply the Liouville field theory approach to minimal conformal
field theory models coupled to two dimensional quantum
gravity [\DDK].
A general $N$-point correlation function, evaluated on a genus $p$
worldsheet is given by
$$
\corr{\prodone N\psi _j}
=\int_{\c M_p}\!\!dm\int\!\!D_{\!\hat g}\phi\,e^{-S_L}\prodone N
e^{\beta(h_j)\phi(z_j)}\corr{\prodone N\opcft j}_{\hat g, p}
\eqno\eq
$$
Here, the Liouville action is given by
\foot{We have included a factor of ${1 /\alpha ^2}$ in $\mu$
which will turn out to be convenient in section 3.}
$$
S_L =\fpi \int\sqrt{\hat g} \left[
	\half \phi \Delta _{\hat g} \phi -\kappa  R_{\hat g} \phi
	+{ \mu \over \alpha ^2}  e^{ \bestar\phi}
   	\right]
\eqno\eq
$$
The measure $D_{\hat g} \phi$ is the translation invariant Lebesgue
measure on $\phi$, which yields a quantization of 2-D
gravity provided $\kappa$ and $\alpha$ are related to the matter
central charge. [\CDG,\DDK,\MDP,\EDH]
$$
3\kappa ^2 = 25-c \qquad \qquad {\rm and} \qquad \qquad
\alpha ^2 + \kappa \alpha +2 =0
\eqno\eq
$$
The gravity-matter vertex operators $\psi _j$ are constructed by
$$
\psi _j = \psi^{\rm matter} _j e^{\beta (h_j) \phi }
\qquad \qquad
\beta(h_j) ={-\sqrt{25-c}+\sqrt{1-c+24h_j}\over2\sqrt3}
\eqno\eq
$$
The requirement that these operators be physical (i.e. dimension (1,1))
determines $\beta (h_j)$ in the above formula [\KPZ,\DDK].
The matter operators $\psi^{\rm matter} _{r_j' r_j}$
are spin zero primary fields
of conformal dimension $h_j$.
The matter correlation function is evaluated for metric $\hat g$ and
genus $p$ within the matter conformal field theory [\DF].

We shall henceforth specialize to minimal models,
not necessarily unitary. In section 3, we shall provide
a proof of the validity of the formal continuation method, which
is complete only for minimal models.
$$
c=1-{6\over q(q+1)}  \qquad \qquad
	h_{r'r}={\left(qr'-(q+1)r\right)^2-1\over4q(q+1)}
\eqno\eq
$$
Here $q$ is a rational number for general minimal models [\BPZ]
and $q=2,3,4,\dots$ for unitary models [\FQS].
We also have
$$
\alpha= \beta (h_{11}) =-\sqrt{2q\over(q+1)}
\qquad \qquad\
\kappa =(2q+1)\sqrt {2 \over {q(q+1)}}
\eqno\eq
$$
and
$$
\beta(h_{r'r})
={-(2q+1)+\left|qr'-(q+1)r\right|\over\sqrt{2q(q+1)}}
\eqn\dimeq
$$

Following [\GL], we split the integration over $\phi$ into a constant
mode and a piece orthogonal to constants on the worldsheet
$$
\int D_{\!\hat g}\phi\,e^{-S_L}\prodone N e^{\beta(h_j)\phi(z_j)} =
\left({\mu\over4\pi \alpha ^2}\right)^s{\Gamma(-s)\over\bestar}
\int D_{\!\hat g}'\phi \ e^{-S'_L}\left(\int\!\!\sqrt{\hat g}\
  e^{\bestar\phi}\right)^s
\prodone N e^{\beta(h_j)\phi(z_j)}
\eqno\eq
$$
Here $D'_{\!\hat g}\phi$ denotes the integration over the modes orthogonal
to the constant mode and
$$
S'_L = \fpi \int\!\sqrt{\hat g} \left[
	\half \phi \Delta _{\hat g} \phi -\kappa  R_{\hat g} \phi
	\right]
\eqno\eq
$$
The variable $s$ is the total scaling dimension
$$
s=-{\kappa \over\alpha }(1-p)-\sum_{j=1}^N{\beta(h_j)\over\alpha}
\eqn\seq
$$
where $p$ is again the genus of the worldsheet.
In general, $s$ is not an integer; for minimal models
it is a rational number.

We shall now restrict our attention to
the three point function on the sphere ($p=0$).
To facilitate the computation, we shall concentrate the curvature
of the sphere at $z=\infty$.
In this case, the computation reduces to evaluating correlation functions
of free fields on the plane with
the flat metric.
We may put the three fields at  $(z_1,z_2,z_3)=(0,1,\infty)$.
Then the correlation function of the scalar field $\phi$ reduces to
$$
\eqalign{
\int D_{\!\hat g}\phi \,e^{-S_L}
& \prodone 3 e^{\beta(h_j)\phi(z_j)} \cr
&=\left({\mu\over4\pi \alpha ^2 }\right)^s{\Gamma(-s)\over\alpha}
  \int\prod_{i=1}^{s}d^2w_i|w_i|^{-2\bestar\beh1}|1-w_i|^{-2\bestar\beh2}
  \prod_{i<j\atop i,j=1}^{s}|w_i-w_j|^{-2\bestar^2}.\cr}
\eqno\eq
$$
As such, the above  formula can only make sense when $s$ is integer.
We will need to continue in $s$.
The general integral of this type has been calculated in [\DF].
Henceforth we shall be using the notation
$$
\Delta(x)\equiv\Gamma(x)/\Gamma(1-x)\qquad \qquad S(x)\equiv
 {\sin \pi x \over \pi}
\eqno\eq
$$
In terms of $\Delta$, we obtain
$$
\eqalign{{\cal J}_{n'n}(a_1',a_2';a_1,a_2; \rho ', \rho ) &\equiv
	{1\over n'!n!}
	\int\prod_{i=1}^{n'}d^2w'_i |w'_i|^{2a_1'}|1-w'_i|^{2a_2'}
    	\int\prod_{i=1}^{n }d^2w _i |w _i|^{2a_1 }|1-w _i|^{2a_2 }
\cr&\quad\times
  	\prod_{i<j\atop i,j=1}^{n'}|w'_i-w'_j|^{4\rho'}
  	\prod_{i<j\atop i,j=1}^{n }|w _i-w _j|^{4\rho }
   	\prod_{i=1}^{n'}\prod_{j=1}^{n}\left|w'_i-w_j\right|^{-4}
\cr&={\pi}^{n+n'} \rho^{-4nn'}
   	\Delta^{-n'}\!(\rho')    \Delta^{-n}\!(\rho)
	\prod_{i=1}^{n'}\Delta(-n+i\rho')\prod_{i=1}^{n}\Delta(i\rho)
\cr&\quad\times
 	\prod_{i=0}^{n'-1}\prod _{j=1}^3 \Delta(1-n+a'_j+i\rho')
 	\prod_{i=0}^{n -1}\prod _{j=1}^3 \Delta(1+a_j+i\rho)
\cr}
\eqn\inteq
$$
Here
$$
\sum a_i'=2n-2-(2n'-2)\rho' \qquad \qquad \sum a_i =2n'-2-(2n-2)\rho
\eqno\eq
$$
Using this integral formula for $(n',n)=(0,s)$, we find
$$
\int D_{\!\hat g}\phi\,e^{-S_L}\prodone 3 e^{\beta(h_j)\phi(z_j)} =
\left({\mu\over4\alpha ^2 }\right)^s{\Gamma(-s)\Gamma(s+1)\over \alpha}
\Delta^{-s}\!(-\rho')
\prod_{(x',x)=(0,0)\atop (r_j',r_j)}
\prod_{i=1}^{s}\Delta(|x'\rho'-x|-i\rho')
\eqn\corrgeq$$
Here, we used the notation $\rho'\equiv\bestar^2/2$.

The three point function on the sphere,
$\corr{\prodone 3 \opcft j}_{\hat g}$, may be computed
using the method of Dotsenko and Fateev [\DF].
An arbitrary primary field in the Kac table $\{\psi^{\rm matter}_{r'r},\
1\leq r'\leq q+1,1\leq r \leq q\}$ may be expressed using
a free field $\varphi$ as\foot{
We normalized the scalar fields $\phi,\varphi$ according to the standard
convention, $\phi(z)\phi(z')=-\ln|z-z'|^2+\c O(1)$.
This is different from the normalization of [\DF]
by a factor of $\sqrt2$.}
$$
\psi^{\rm matter}_{r'r} =e^{i\al{r'r}\varphi},\qquad
\al{r'r}\equiv\half\left[(1-r')\al-+(1-r)\al+\right],\qquad
\al-=-{2\over\al+}=\alpha
\eqno\eq
$$
with the action
$$
S_\varphi=\fpi \int\sqrt{\hat g}
\left[\half \varphi \Delta \varphi +i\alpha _0  R_{\hat g}
\varphi+e^{i\al-\varphi}+e^{i\al+\varphi}
\right]
\eqno\eq
$$
We shall put the background charge $2\alpha _0 \equiv\al++\al-$ at infinity.
We note in passing that this representation of the primary
field is two--fold degenerate, a fact we shall utilize later.
To satisfy charge conservation including the charge at infinity,
we need to put in ``screening charges", $\int d^2z\,\exp(i\al\pm\varphi)$,
which have conformal dimension $(0,0)$.
Then the three point function is
$$
\eqalign{
\corr{\prodone 3 \opcft j}_{\!\!\hat g}
&=\int D_{\!\hat g}\varphi e^{-S_\varphi}\prodone3 e^{i\alr j\varphi(z_j)}
{1\over n'!n!}\left(\int d^2w'e^{i\al-\varphi(w')}\right)^{n'}
\left(\int d^2w e^{i\al+\varphi(w)}\right)^{n}
\cr
&=
{\cal J}_{n'n}(\al-\alr1,\al-\alr2;\al+\alr1,\al+\alr2;\rho',\rho)
\cr}
\eqno\eq
$$
Here,
$$
2n'\equiv\sum_{j=1}^3r'_j-1,\quad
2n \equiv\sum_{j=1}^3r _j-1,\quad
\rho'=\al-^2/2,\quad\rho=\al+^2/2
\eqno\eq
$$
Using the integral formula \inteq\ we obtain
$$
\eqalign{
\corr{\prodone 3 \opcft j}_{\hat g}
&=\pi^{n+n'}\rho^{-4nn'}
\Delta^{-n'}\!(\rho') \Delta^{-n}\!(\rho)
  \cr&\qquad\times
\prod_{(x',x)=(0,0)\atop (r_j',r_j)}
\prod_{i=1}^{n'}\Delta(x-n+(-x'+i)\rho')
\prod_{i=1}^{n}\Delta(x'+(-x+i)\rho)\cr}
\eqn\corrcfteq
$$

This form has the advantage that it is manifestly symmetric under
the interchange of three fields, unlike the formula obtained in [\DF].
It differs from their formula by a normalization factor associated
with the three fields.
However, it suffers from the ambiguity of the type 0/0
when $r',r$ are integers.
These ambiguities only exist in factors which may be
absorbed in the normalization of the external fields.
Furthermore, when we combine
the matter part with the gravitational part, these
ambiguities cancel out. We may take $r',r$ to be integer
when we have combined these factors.

Combining the gravitational part of the correlation function,
\corrgeq, and the conformal field theory part, \corrcfteq,
 the three point function of the minimal models
coupled to two dimensional gravity is
$$
\eqalign{
	\corr{\prodone 3 \psi_j}
&=\bigl ({\mu \over 4\alpha ^2}\bigr )^s
	\pi^{n+n'}\rho^{-4nn'-2s}\bestar^{-1}
\Delta^{-n'+s}\!(\rho') \Delta^{-n}\!(\rho)
\Gamma(-s)\Gamma(s+1)
  \cr&\qquad\times\!\!
\prod_{(x',x)=(0,0)\atop (r_j',r_j)}
\prod_{i=1}^{s}\Delta(|x-x'\rho'|-i\rho')
\prod_{i=1}^{n'}\Delta(x-x'\rho'-n+i\rho')
\prod_{i=1}^{n}\Delta(x'-x\rho+i\rho)\cr}
  \eqn\correq$$
We shall now reduce the product over $s$ factors, using the
following basic rearrangement formula
$$
\prod _{k=0} ^{n-1} \Delta ({x \over n} + k {m \over n})
=({m\over n}) ^{2x-m-n+mn} \prod _{l=0} ^{m-1} \Delta ({x\over m}
+l{n\over m})
\eqn\mult
$$

In a minimal model, two operators in the Kac table
represent the same field.
Namely, $\psi^{\rm matter}_{r',r}$
and $\psi^{\rm matter}_{q+1-r',q-r}$ represent the same field.
The three point functions as computed above obey fusion
rules which are {\it not} invariant under the reflection $(r',r)
\leftrightarrow(q+1-r',q-r)$ when applied to one of the three fields.
However, the three point function {\it is} invariant when the
above reflection is applied to two fields simultaneously
(up to the normalization of the fields).
Since $(r'\rho'-r) \leftrightarrow -(r'\rho'-r)  $
under this operation, $ r_j-r'_j\rho'$ may be assumed to be
all of the same
sign for the three point function
without any loss in generality. We shall treat the two cases
when they are all positive and all negative separately.

When $r_j-r'_j\rho'\geq0$ for $j=1,2,3$, from \dimeq\ and \seq, we obtain
$$
\rho'={n\over n'+s+1}
\eqno\eq
$$
Using the basic rearrangement formula \mult\
the following formula holds in this case:
$$
\prod_{i=1}^s\Delta(y-i\rho') =
   \rho^{(n+1-2y)(n'+s+1)-n}\Delta^{-1}\!(y)
   \prod_{i=1}^n\Delta^{-1}\!((-y+i)\rho)
   \prod_{i=1}^{n'}\Delta^{-1}\!(y-n+i\rho')
\eqn\rearr
$$
Using this formula for $y=0,r_j-r'_j\rho'$ and
from \correq, we obtain the formula for the three point function as
$$
\corr{\prodone 3 \psi_j}
=\bigl ({\mu \over 4 \alpha ^2}\bigr ) ^s
\pi^{n+n'}\rho^{2n'-2n+2}\bestar^{-1}
\Delta^{-n'+s}\!(\rho') \Delta^{-n}\!(\rho)
\prodone3 \Delta^{-1}(r_j-r'_j\rho')
\eqno\eq
$$
We have set a factor of $\Gamma(-s)\Gamma(s+1)S(s)$
which is of the form $0/0$ when $s$
is integer to be one. This formal procedure
is justified by establishing the identity for
asymptotic values of $s$ in the semi--classical
analysis of the next section.
We rescale   the external fields as
$$
\psi _j\mapsto\pi^{(r_j+r'_j)/2} (4\alpha ^2)^{(r'_j-r_j\rho)/2}
\rho^{(r'_j-r_j)/2} \Delta^{r_j\rho/2-r'_j}\!(\rho')
\Delta^{-r_j/2}\!(\rho)
\Delta^{-1}\!(r_j-r'_j\rho')\psi_j
\eqn\resc
$$
This rescaling is not singular. To see this, note
that neither $\rho$ nor $\rho'$ are integer valued so that the first
two factors of $\Delta$ are neither zero nor infinity.
Next, $r_j-r'_j\rho'$ is never integer when $r_j$ and
$r'_j$ belong to the Kac table of a minimal $(p,p')$ model
with $\rho=p/p'$ where $p,p'$ are relatively prime.
This follows from the fact that $1\leq r_j\leq p'-1$ and
$1\leq r'_j\leq p-1$.
Similarly, $r'_j-r_j\rho$ is never integer.
We furthermore  rescale  by factors that are independent
of the external indices $(r'_j,r_j)$,
the three point function reduces just to
$$
\corr{\prodone 3 \psi_j} = \mu ^s
\eqn\final
$$

Similarly,
when $r_j-r'_j\rho'\geq0$, we obtain
$$
\rho'={n+1\over n'-s}
\eqno\eq
$$
The  following formula can be shown to hold in this case using \mult\ again
$$
\prod_{i=1}^s\Delta(-y-i\rho') =
   \rho^{(n-2y)(n'-s)-n-1}\Delta^{-1}\!(-y\rho)
   \prod_{i=1}^n\Delta^{-1}\!((-y+i)\rho)
   \prod_{i=1}^{n'}\Delta^{-1}\!(y-n+i\rho')
\eqn\rearrinv
$$
Using this formula for $y=0,r_j-r'_j\rho'$,
the three point function reduces to
$$
\corr{\prodone 3 \op_j}
=\bigl ({\mu \over 4 \alpha ^2}\bigr )^s
\pi^{n+n'}\rho^{2n'-2n-2}\bestar^{-1}
\Delta^{-n'+s}\!(\rho') \Delta^{-n}\!(\rho)
\prodone3 \Delta^{-1}(r'_j-r_j\rho)
\eqno\eq
$$
The same rescaling as before with $\Delta(r_j-r'_j\rho')$
replaced by $\Delta(r'_j-r_j\rho)$ again reduces the correlation function
to $\mu^s$ of \final.

Differentiating with respect to $\mu$ is equivalent to
bringing down the area operator:
$$
{\d\over\d\mu}\corr{\prodone N\op_j}
=\corr{{\bf 1}\prodone N\op_j}
\eqno\eq
$$
The three point function which is independent of the normalization
of the fields may be computed using this relation as
$$
{\corr{\op_1\op_2\op_3}^2{\rm Z}\over
\corr{\op_1\op_1}\corr{\op_2\op_2}\corr{\op_3\op_3}}
 ={\prodone3\left|qr'_j-(q+1)r_j\right|\over(q+1)(2q+1)}
\eqno\eq
$$
where ${\rm Z}$ is the partition function of the model on the sphere.
This agrees with the three point function computed in
the hermitian matrix model [\DK].

\chapter{Formal Continuation in the Number of Liouville Screening Charges}

Does the formal continuation procedure in the number of Liouville
screening operators $s$ make sense?
In the derivation of [\GL], and again
in the preceding section, an expression defined only for integer $s$
like formula \corrgeq, was manipulated in such a way as to obtain an
expression that would make sense for all complex values of $s$, such as
in formula \rearr.  This type of manipulation may not in general be
permitted, as it is inherently ambiguous up to periodic functions.

In this section, we shall compare the final formula gotten by
formal continuation directly with the Liouville functional
integral.  We know already that these two expressions agree on all
positive integer values of $s$.
Furthermore, we clearly have a meromorphic function of $s$.
Thus, in order to show that the formal continued formula indeed
yields the answer for the functional integral for all $s$, it is
necessary to verify that the asymptotic behavior as $|s|
\rightarrow \infty$ matches.

A meromorphic function that vanishes on all positive integers, and also
at $\infty$ must be identically zero.  Using an analogous result,
the agreement on
the integers and the asymptotics, we will have shown the validity of
the final answer, and thus of the analytic continuation procedure.

The significance of the large $|s|$ limit is that of a weak coupling
expansion.
To see this, consider \seq\  expressing $\beta$ through \dimeq
$$
s = {1 \over \alpha^2} \biggl \{ 2 + \alpha^2 + \sum_{j=1}^3
\biggl ( \bigg | r_j - {\alpha^2 \over 2} r_j' \bigg | - {\alpha^2 \over
2} - 1 \biggr ) \biggr \},\qquad
s  = {2n \over \alpha^2} - n'-1
\eqno\eq
$$
where the last identity holds when
$r_j$ and $r'_j$ are fixed and $|s| $ is sufficiently large.
Thus, the limit we shall consider corresponds to holding $r_j$ and
$r_j'$ fixed, but letting $\alpha \rightarrow 0$ or $c
\rightarrow - \infty$.
{}From a field theoretic point of view, this
is the semiclassical limit with $\alpha^2 \sim \hbar$.  The functional
integral may now be recast in a form that exhibits the full
$\alpha^2$-dependence
$$
I = \int D_\g \phi ~e^{- {1 \over \alpha^2}S_o(\phi)-S_1(\phi)}
\eqno\eq
$$
where the $\alpha^2$-independent actions are given by
$$
S_o(\phi) = {1 \over 4\pi} \int\! \sqrt \g ~\biggl[ \half \phi \Delta_ \g
\phi + 2 R_\g \phi + \mu e^\phi
+ 4\pi \sum_{j=1}^3 \phi(z) (r_j -1) \delta_\g (z, z_i) \biggr ]
\eqno\eq
$$
and
$$
S_1(\phi) = {1 \over 4\pi} \int\! \sqrt \g ~R_\g \phi -  \half
\sum_{j=1}^3 \phi(z_j) (r_j' +1).
\eqno\eq
$$
The small $\alpha^2$-expansion proceeds by identifying the saddle point
$\phi_o$
of $S_o$, evaluating its classical action, and then expanding in a
perturbative series.
The final form of the answer is
$$
I = e^{-{a_o / \alpha^2}} \pmatrix {1 \over \alpha^2}^\nu (a_1 + a_2
\alpha^2 + a_3 \alpha ^4 + \dots )
\eqn\asym
$$
Here $a_o $ is essentially $ S_o (\phi_o)$, a real number dependent only on the
$r_j$'s, $2\nu$ is the number of normalizable zero modes of the operator
$$
{\delta ^2 S_o \over \delta \phi(x) \delta \phi(y)} \bigg |_{\phi =
\phi_o}
\eqn\opeq
$$
and the coefficients $a_i$ result from the contribution with $i$-loops.

We begin by constructing the instanton solution
\foot{Expansions around classical solutions were studied in
[\EFJ,\EDH]}
$\phi_o$.  It is prudent
to work with the round metric on the sphere, where
$$
\gm = e^\sigma \delta _{mn} \qquad \qquad \sigma = -\ell n~ \half~
(1+|z|^2)^2
\eqno\eq
$$
The saddle point equation
$$
\Delta _\g \phi_o + 2 R_\g + \mu ~e^{\phi_o} = - 4\pi \sum_{j=1}^3
(r_j -1) \delta _\g (z, z_j)
\eqn\liou
$$
admits regular real solutions for $\mu = - |\mu| < 0$, which is the
case we shall study:
$$
\phi_o(z) = - \ell n~ {|\mu| \over 2} ~ {(1+ |A(z)|^2)^2 \over (|A'(z)|^2
+ \epsilon^2) (1+|z|^2)^2}
\eqno\eq
$$
Here $A(z)$ is a meromorphic function such that $A'(z)$ has a zero of
order $(r_i-1)$ at $z_i$ and $\epsilon$ is an infinitesimal positive
constant.
For $\phi_o(z)$ to be regular at $z = \infty$
(recall the only singularities occur at $z_i$'s), we should have
$$
A(z) \sim a z \qquad {\rm as} \qquad |z| \rightarrow \infty \qquad a
\not =0
\eqno\eq
$$
Hence $A$ is the ratio of two polynomials $P,\ Q$
$$
A(z) = {Q(z) \over P(z)} \qquad \qquad \deg Q = \deg P + 1
\eqno\eq
$$
To determine the total number of zeros in $A'$ requires some extra care.
There are $r_1 + r_2 + r_3 - 3 = 2n -2$ zeros from the $z_i$'s, which
produce the $\delta$-functions on the right hand side of \liou.
However,
recall that there is also a charge at $z = \infty$, due to the curvature
term.  This is perhaps most easily seen when considering the flat
metric on the plane plus a curvature $\delta$-function with charge 2 as
$z = \infty$ to make the plane into a sphere.
The total number of zeros
of $A'$ is then $2n$, hence $\deg Q = n + 1$.
We normalize $P$ such that
$$
P(z) = \prod_{k=1}^n (z - w_k)
\eqno\eq
$$
and of course, none of the $w_k$'s should equal $z_i$.

{}From the requirement that only double poles should occur in the
expansion of $A'(z)$, we get an equation determining $w_k$
\footnote\dagger{The point at $\infty$ disappears in this equation!}
$$
\sum_{i=1}^3 {r_i-1 \over w_k-z_i} = 2 \sum _{\ell = 1\atop \ell \not=
k} ^n {1 \over w_k-w_\ell} \qquad 1 \leq k \leq n.
\eqn\chargeeq
$$
This equation governs 3 fixed charges at $z_i$ with strength $- {r_i-1
\over 2}$ and $n$ charges of unit strength that settle into an
equilibrium configuration under the forces of 2-D electrostatics.  The
equation \chargeeq\  is invariant under complex M\"obius transformations.

Next, we evaluate the classical action on this saddle point $\phi_o$,
and we find
$$
\eqalign{ S_o(\phi_o) & = - 4n - 2n ~\ell n |\mu| + 2n ~\ell n ~2
 	- 2 \sum_{k=1}^ n  \ell n~ {|Q(w_k)|^2 \over |a|^2}\cr
& 	+ \half \sum_{i,j=1}^3 (r_i-1) (r_j-1)~ \ell n ~(|z_i-z_j|^2 +
	\epsilon^2)
 	+ 2 \sum _{i=1}^3 (r_i-1) ~\ell n~ (1+|z_i|^2)\cr}
\eqno\eq
$$
Notice that this expression is independent of the normalization factor
$|a|$.  As it stands, this answer is infinite as $\epsilon
\rightarrow 0$, but of course, we should keep in mind that the
self-energies of the charges $z_i$ must be subtracted.
Furthermore, $e^{-{1 \over \alpha^2}S_o(\phi_o)}$ is a three point
function in a conformal field theory, so its dependence on the points
$z_i$ is fixed by conformal invariance.
We have
$$
e^{-{1 \over \alpha^2}S_o(\phi_o)} = e^{-{ a_o \over \alpha^2}} \prod_{i<j}
|z_i - z_j|^{2h_i+ 2h_j - 2h_k-2}\prod _i (g_{z\bar z})^{h_i}
\eqno\eq
$$
we obtain the following final expression for $a_o$:
$$
a_o =-4n - 2n ~\ell n ~|\mu| + 2n ~\ell n ~2 - \ell n |F|^2
\eqno\eq
$$
with the conformal invariant $F$ given by
$$
F = \prod _{k=1}^n {Q(w_k)^2 \over a^2} \prod _{i<j} (z_i-z_j)^
{-(r_i-1) (r_j-1) - \half (r_i^2-1) -\half (r_j^2-1) + \half (r_k^2 -1)}
\eqno\eq
$$
The quantity $Q(w_k)$ is also easily calculated from the fact that
$$
(Q'P-QP')(z) = a \prod_{i=1}^3 (z-z_i)^{r_i-1}
\eqno\eq
$$
so we find
$$
Q(w_k)^2 = a^2(P'(w_k))^{-2} \prod _{i=1}^3 (w_k - z_i)^{2r_i-2}
\eqno\eq
$$
Using conformal invariance of $F$, we let $z_1 = 1,z_2 = -1, z_3 =
\infty$.
$$
F =2^{2n^2 + 2n - 2n {(r_1+r_2)}}
	 P(1)^{2r_1-2} P(-1)^{2r_2-2} \prod_{k=1}^n P'(w_k)^{-2}
\eqn\feq
$$

Next, we have to evaluate the solutions $w_k$ of \chargeeq\ or equivalently
the polynomial $P$.
To do so, we slightly generalize a clever trick due
to Szego [\SZE].
The Jacobi polynomials $y = P_n^{(\alpha,\beta)} (x)$
satisfy the following differential equation
$$
(1-x^2)y'' + (\beta - \alpha - (\alpha + \beta + 2)x) y' + n (n + \alpha
+ \beta + 1) y = 0.
\eqno\eq
$$
Hence, at a zero $w_k$ of $ P_n^{(\alpha, \beta)}$ for $ 1 \leq k \leq n$
and with $w_k \not= \pm~ 1$  we have
$$
{P_n^{''(\alpha,\beta)}(w_k) \over P_n^{'(\alpha, \beta)} (w_k)}
= - {1+\alpha \over w_k-1} - {1+\beta \over w_k+1}
\eqn\peq
$$
On the other hand, using the product formula for $P_n^{(\alpha, \beta)}$
in terms of its zeros
$$
{P_n^{''(\alpha,\beta)}(w_k) \over P_n^{'(\alpha, \beta)} (w_k)}
=  2 \sum _{\ell = 1\atop \ell \not= k} ^n {1 \over w_k-w_\ell}.
\eqn\ppeq
$$
For $\alpha = - r_1$ and $\beta = - r_2$, \peq -\ppeq\ precisely coincides
with \chargeeq\ in which, using conformal invariance, we have moved $z_1,
z_2, z_3$ to $1, -1, \infty$.
Thus, the solutions to \chargeeq\ are the
zeros of Jacobi polynomials $w_k$, and $P(w)$ is proportional to
$P_n^{(-r_1,-r_2)}(w)$.

There is a point that requires clarification here :  the Jacobi
polynomials $P_n^{(-r_1, -r_2)}(w)$ can in general have zeros at $\pm
1$, when $r_1$ and $r_2$ are positive integers.
In this case, the above reasoning would not be valid.
Thus we may always take $n$ integer, but
assume that $r_i$'s are slightly away from their integer values.
The final answer for $a_o$ will in fact admit a limit as $r_i$ tends to
integer.

It remains to evaluate \feq.
Fortunately, this problem was also
already solved by Szego [\SZE], and the final answer is
$$
\ell n ~F = \sum_{k=1}^n ~~\sum_{x=0, r_1, r_2, r_3} (x-k) \ell n
|x-k|^2
\eqno\eq
$$
so that
$$
a_o = -4n - 2n ~\ell n |\mu| + 2n~\ell n ~2 - 2 \sum_{k=1}^n~~ \sum
_{x=0,r_i} (x-k) \ell n |x-k|^2
\eqno\eq
$$

Next, we need to determine the number of zero modes of the operator
\opeq, which governs the small fluctuation problem
$$
{1 \over \alpha ^2} S_o (\phi _o + \alpha \varphi )
={1 \over \alpha ^2} S_o (\phi _o)
+\fpi \int \! \sqrt {\hat g} \biggl [ \half \varphi \Delta _{\hat g}
\varphi - \half |\mu | e^{\phi _0} \varphi ^2 \biggr ] +{\cal O}(\alpha)
\eqno\eq
$$
The small fluctuation problem has zero modes determined by the equation
$$
\Delta_\gamma \varphi_o - 2  \varphi_o = 0
\qquad\qquad
\gamma_{mn} = {2|A'|^2 \over (1+|A|^2)^2} \delta_{mn}
\eqn\sphereeq
$$
The metric $\gamma$ is a round metric on the $n$-covered sphere
with curvature $1.$
Thus \sphereeq\ is an eigenvalue equation for the Laplacian on the sphere
(i.e.
$\vec L ^2 $)
with eigenvalue $j(j+1) = 2$.
Hence $j=1$, and there are 3 normalizable zero modes.
Thus $2\nu = 3$ in \asym .
The $1$-loop
contribution is easy to compute as well; the small fluctuation problem
just yields the determinant of
$\vec L ^2 $
on the sphere,
which is a number independent of $r_i$'s.
So the only relevant term comes from evaluating $S_1(\phi_o)$.
It is easy to see that this corresponds to letting
$$
{2n \over \alpha^2} \rightarrow s = {2n \over \alpha^2} +
2n',\qquad
 {1-r_i \over \alpha} \rightarrow \beta_i = {1-r_i \over \alpha }
+ {\alpha \over 2} (1+r_i').
\eqno\eq
$$
We shall not need the explicit expression for $a_1$ here.

It remains to evaluate the asymptotics of the right hand side of \corrgeq,
\rearr\ and \rearrinv.
Actually, here it is convenient to recast the formal continuation formula
in a form closer to that obtained originally in [\GL]
$$
\eqalign{&\int D\phi e^{-S_L}\prod_{j=1}^3 e^{\beh j \phi(z_j)}=
(-1)^{n'+1}\left(-\mu\over4\alpha^2\right)^s
\inv\alpha
\Delta^s(1+\rho')\rho^{2(2n'n+n'-n+s-1)}
\cr&\qquad\times
{\prod_{i=1}^{n'} S(i\rho')\Gamma^2(1+n-i\rho')\over
\prod_{i=1}^{n-1} S(i\rho) \prod_{i=1}^{n} \Gamma^2(i\rho)}
\cr&\qquad\times
\prod_{j=1}^3\inv\Delta\!(r_j-r'_j\rho')\,
{\prod_{i\neq r'_j\atop1}^{n'}S\left((r'_j-i)\rho'\right)
\prod_{i=1}^{n'}\Gamma^2(1+n-r_j+(r'_j-i)\rho')\over
\prod_{i\neq r_j\atop1}^{n}S\left((r_j-i)\rho\right)
\prod_{i=1}^{n}\Gamma^2(r'_j-(r_j-i)\rho)}\cr} \eqn\gl
$$
There are various potential contributions in this formula that are not
of the form \asym.  In each $\Gamma$ and $S$-function there are factors
of the type $s^s$ and $e^{irs}$ with $r$ real.  It is straightforward to
check that all factors form $s^s$  cancel, as well as those of the form
$e^{irs}$, keeping in mind that $\mu < 0$.  With a little patience, we
can evaluate the leading exponential and power asymptotics of the right
hand side of \gl.
$$
I = s^{3/2}~\mu^s
 \exp \biggl \{ 2s + {2s \over n} \sum _{k=1}~~\sum_{x=0, r_1, r_2,
r_3} (x-k) \ell n |x-k| \biggr \} \times {\cal O}(s^0)
\eqn\asympp
$$
This formula precisely agrees with our semiclassical result.
Since the \c O$(s^0)$ term agrees for integer $s$, it agrees
for all $s$. We
conclude that the asymptotics of both sides of equation \gl\ match.

Any zeros and poles on the right hand side of
\gl\ can only occur for $\rho$ rational, and when
$\Re(s)>0$, only for $\rho>0$.
This means that we are dealing with a minimal model with
$\rho=p/p'$, $p,p'$ relatively prime.
If $1\leq r_j\leq p'-1$ and $1\leq r_j'\leq p-1$, \ie\
$(r'_j,r_j)$ is in the Kac table of the minimal model, then the right
hand side of \gl\ has neither zeros nor poles.
To see this, notice in the last line of \gl\ the first
factor was already shown to be neither equal to zero nor infinity
in a discussion after formula \resc; whereas the second and
third factors are regular by an argument similar
to the one after \resc. The factors on the first and
second lines in \gl\ have zeros and poles, but
these only occur for $\Re(s)<0$.
Operators corresponding to $(r'_j,r_j)$ outside the
Kac table, may be related by OPE with known singularities
to operators within the Kac table plus screening operators.
For $(r'_j,r_j)$ outside the Kac table, this accounts for the
zeros and poles on the right hand side of \gl.

On the left hand side of \gl, we have a well--defined Euclidean
functional integral for all values $\Re(s)>0$ provided $(r'_j,r_j)$
belongs to the Kac table and no singularities occur as
a function of coupling constant.
Outside the Kac table, new ultraviolet divergences occur governed
by the OPE discussed above.

To conclude the argument, we divide both sides by the
asymptotic $|s|\rightarrow\infty$ behavior of \asympp,
so that both tend to a constant as $|s|\rightarrow\infty$.
Since both sides agree on $s$ integer, these constants
are the same.

The difference of both sides is now a meromorphic function
for $\Re(s)>0$, tends to zero as $|s|\rightarrow\infty$
and vanishes for integer $s$.
In addition, we have shown above that for $(r'_j,r_j)$ inside
the
Kac table, both sides have no poles in $\Re(s)>0$ and hence are
holomorphic in $\Re(s)>0$.
For $(r'_j,r_j)$ outside the Kac
table, we argued that the singularities on both sides are identical,
so that the difference is again holomorphic when $\Re(s)>0$.
By a standard theorem of complex analysis [\BOAS], such a function
vanishes identically and hence we have
shown the validity of \gl.

\chapter{Summary}

We have shown that the formal continuation procedure proposed in
[\GL] in the case of the $3$--point function holds in
the case of minimal models,
directly from semiclassical
methods on the Liouville functional integral.
This increases our confidence in the procedure considerably,
and puts the free field,
screening operator approach to Liouville theory on a footing close to
equal with the Dotsenko-Fateev approach to calculating correlation
functions in minimal models.

We generalized the calculation of [\GL,\KIT] to the
case of arbitrary external physical states in minimal models using just
Liouville theory.
We obtain the same answer as in [\DOT] who uses two
screening operator instead of the Liouville theory.
It would be very interesting to generalize this approach to compute
$N$-point correlation functions on the sphere and on the torus.
The generalization to supergravity is in a companion paper [\SUP].

\bigskip
\noindent\undertext{Acknowledgments}\nl
One of us (E.D) acknowledges  useful discussions with Jean-Loup
Gervais, Mark Goulian
and David Gross and the hospitality of the Aspen Center for Physics
where part of this work was carried out.
\refout
\endpage\end